\documentclass[%
 aps,
 jmp,%
 amsmath,amssymb,
 reprint,%
]{revtex4-1}

\usepackage{graphicx}
\usepackage{dcolumn}
\usepackage{bm}

\begin{document}


\title{The underdamped Brownian duet and
stochastic linear irreversible thermodynamics}

\author{Karel Proesmans}
\author{Christian Van den Broeck}%
 
 \affiliation{Hasselt University, B-3590 Diepenbeek, Belgium.}
 
\date{\today}

\begin{abstract}
Building on our earlier work (Proesmans et. al., Phys.~Rev.~X \textbf{6} (2016), 041010), we introduce the underdamped Brownian duet as a prototype model of a dissipative system or of a work-to-work engine. Several recent advances from the theory of stochastic thermodynamics are illustrated with explicit analytic calculations and corresponding Langevin simulations. In particular, we discuss Onsager-Casimir symmetry, the trade-off relations between power, efficiency and dissipation, and stochastic efficiency. 
\end{abstract}

\maketitle
\section{Introduction}
The theory of linear irreversible thermodynamics is typically introduced as a special topic in an advanced class on thermodynamics. 
It usually focuses on the derivation of Onsager symmetry and its application to thermo-electric effects, while 
the Prigogine minimum entropy production theorem is occasionally included. The derivation of Onsager symmetry itself is often
clouded in the somewhat vague Onsager regression hypothesis, stating that fluctuations on average regress in the same way as 
externally produced perturbations. The discussion of this issue is actually quite subtle. A related jump from statements about the
 micro world to macro world concerns the validity of a  microscopic derivation for linear response or Green Kubo relations.
 There are several other concerns: the Onsager symmetry is by no means general. The most obvious generalization is the
 Onsager-Casimir symmetry, where one needs to distinguish between time-symmetric quantities (such as position)
  and time-antisymmetric variables (such as velocities and magnetic field). Furthermore, a proper definition of thermodynamic
  forces and fluxes is required to get a bona fide thermodynamic description including the bilinear law for the corresponding entropy production. 
  Another usual  gap  in the whole presentation is the lack of the connection with one of the founding principles of thermodynamics, namely that of the thermodynamic engine.
  
 The purpose of this paper is to address all of these issues, by introducing a simple exactly solvable model.
  It consists of a particle in a harmonic potential subject to a time-periodic force. The time-periodicity can be linked to the time-periodic operation of 
 most thermodynamic engines. The full dynamic and thermodynamic description, including the first and the second law, the Onsager coefficient(s) and the thermodynamic efficiency of the related engine, can be derived via a simple explicit calculation  without any extraneous assumptions. 
 The Onsager coefficients display the Onsager-Casimir symmetry, including the time-reversal of the periodic driving (which needs not be time-symmetric). 
  
The additional purpose is to present the recent spectacular advances in our understanding of the second law by considering its application to small scale systems.
Hence we revisit the above scenario for a  {\bf Brownian} particle, i.e.,  a particle which is small enough to be subject to the thermal fluctuations. Assuming a description in 
 terms of a Langevin equation  with the usual additive Gaussian white noise, the above dynamic and thermodynamic discussion can be repeated.  This analysis is
 the generalization to the underdamped case of the Brownian duet considered in \cite{proesmans2016brownian}.  The Langevin description incorporates the property of detailed balance, which reflects the micro-reversibility of the underlying dynamics. We show that  the implied fluctuation dissipation response relations are equivalent with the fluctuation theorem, which is the generalization of the second law to small systems.  We discuss the implications for the stochastic efficiency of the engine, and show that they are fully described in terms of the afore derived Onsager coefficients.  We illustrate all the properties by Langevin simulations. They are, as expected in the presence of exact analytic results, in full agreement with the theory. We in particular illustrate several of the surprising findings  in this context, notably that the reversible efficiency is the least likely in the long time limit for engines operating under  time-symmetric driving.

\section{Underdamped particle In a harmonic potential}

Consider a particle with mass $m$, moving in a one-dimensional harmonic potential with spring constant $\kappa$, subject to an external time dependent force $F(t)=F_0 g(t)$ and a friction force proportional to the speed with friction coefficient $\gamma$. We will refer to the particle as being the system, while its surrounding responsible for the friction force is supposed to be a thermal reservoir at temperature $T$.
The Newton equation of motion reads:
\begin{equation}\label{meq}
    m\ddot{Z}(t)=-\gamma \dot{Z}(t)-\kappa Z(t)+F(t).
\end{equation}
For long enough times, the dependence on the initial position is forgotten and one can concentrate on the following "steady state" time-dependent solution  $Z(t)$ of this equation:
\begin{equation}\label{Zeq}
    Z(t)=\frac{2F_0}{\kappa\tau_2}\int_0^{\infty}dt' e^{-\frac{t'}{2\tau_1}}\frac{\sinh\left({\sqrt{1-4\tau_1/\tau_2}}\frac{t'}{2\tau_1}\right)}{\sqrt{1-4\tau_1/\tau_2}}g(t-t').
\end{equation}
Here, we identified three intrinsic time-scales of the damped harmonic oscillator:  the relaxation time in the absence of a spring, $\tau_1=m/\gamma$, the overdamped relaxation time in the "absence of a mass"  $\tau_2=\gamma/\kappa$, and the oscillation period in the absence of friction $\tau_3=2\pi \sqrt{m/\kappa}=2\pi \sqrt{\tau_1 \tau_2}$.  Furthermore, the transition from the underdamped to the overdamped regime is described by the critical ratio $\tau_2/\tau_1=4$. 

\begin{figure}
\center
\includegraphics[width=0.45\textwidth]{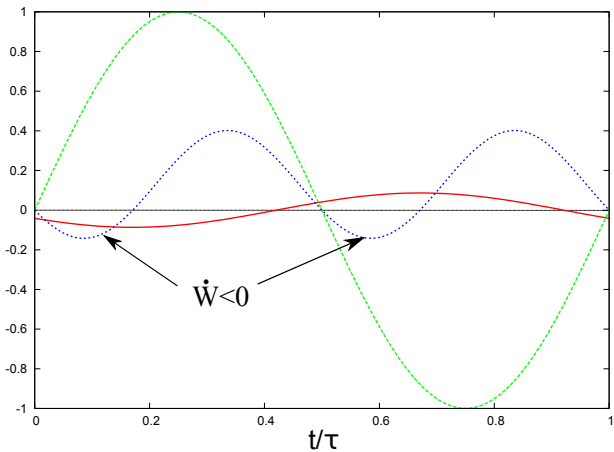} 
\caption{Scaled position $10Z(t)\kappa/F_0$ (full line), force $F(t)/F_0$ (dashed line) and power $10\dot{W}\tau\kappa/(F_0^2)$ (dotted line) of a bead with $F(t)=F_0\sin(2\pi t/\tau)$, in function of time for $\tau_1=\tau_2=\tau$. Note that while $\bar{\dot{W}}\geq0$, the system transiently returns part of its energy back to the worksource (cf. the two time-intervals where $\dot{W}\leq0$).} 
\label{fig1}     
\end{figure}

We will be particularly interested in the case of  a time-periodic forcing $F(t)=F(t+\tau)$. It follows that $Z(t)$ is also periodic with the same period. In the case of a sine modulation $F(t)=F_0\sin2\pi t/\tau$, one finds, see also
 Fig.~\ref{fig1}: 
\begin{equation}
Z(t)=\frac{F_0}{\kappa}\frac{\left(1-4\pi^2\alpha_1\alpha_2\right)\sin\left({2\pi \alpha}\right)-2\pi\alpha_2\cos\left({2\pi \alpha}\right)}{\left(1-4\pi^2\alpha_1\alpha_2\right)^2+4\pi^2\alpha_2^2},
\end{equation}
with \begin{equation}
\alpha_1=\tau_1/\tau,\;\;\;\alpha_2=\tau_2/\tau,\;\;\; \alpha=t/\tau
\end{equation}

Having solved the dynamics of the problem, we turn to  its thermodynamics. 
The first law states the conservation of total energy. The particle energy is given by 
\begin{equation}E=mV^2/2+\kappa Z^2/2,\end{equation} 
with $V=\dot{Z}$, while the power exerted by the external force is given by $\dot{W}=F\dot{Z}$. The notation of power as $\dot{W}$ should not be misinterpreted:  the latter is not a full time-derivative, corresponding to the well known fact that there no such thing as a state variable work $W$.
By multiplying the equation of motion Eq.~(\ref{meq})  with $\dot{Z}$, one immediately deduces the following balance equation:
\begin{equation}\label{fleq}
    \dot{E}=\dot{W}+\dot{Q},
\end{equation}
via which we identify the rate of heat (defined as heat towards the system, i.e., away from the reservoir):
\begin{equation}\label{qeq}
 \dot{Q}=-\gamma \dot{Z}^2.
\end{equation}
We recognize the familiar expression of the Joule heating rate $- \dot{Q}=\gamma \dot{Z}^2\geq0$, being the heat flux dumped into the reservoir. Again, one should beware of the notation $\dot{Q}$, as this does, in general, not represent the time derivative of a quantity $Q$ (see however comment \footnote{In the present context, this statement is not correct: a thermal reservoir is a system at equilibrium which exchanges energy with the external work exclusively via heat. Hence the latter is equal to the change of internal energy, and is de facto a state variable.}).

Having identified the heat flux, one can turn to the second law of thermodynamics and the entropy production. 
We are using here the formulation of thermodynamics for open systems as introduced by Prigogine. The entropy change of a system is the sum of two contributions: the irreversible entropy production $\dot{S}_i$ and the entropy exchange $\dot{S}_e$ with the environment:
\begin{equation}\label{epb}
 \dot{S}=\dot{S}_i+\dot{S}_e.
 \end{equation}
The entropy flow is given in terms of the heat flux, while the entropy production is nonnegative:
\begin{eqnarray}
 \dot{S}_e=\frac{\dot{Q}}{T}\;\;\;\;\ \dot{S}_i\geq 0.
 \end{eqnarray}
  
With the application to thermodynamic engines in mind, we again focus on the case of a time-periodic forcing. 
 $Z(t)$ is then also periodic with the same period, and hence so are all other the thermodynamic quantities,  in particular ${E}$ and $ {S}$. 
The obvious thing to do is to investigate the averages of these quantities over one period. Such an average will be designated by an overbar: $\bar{y}=\int_0^{\tau} dt\; y(t)/\tau$.
Since the energy returns to its original value after each period, one has:
\begin{equation}\label{flpeq}
\bar{\dot{E}}=\bar{\dot{Q}}+\bar{\dot{W}}=0.
\end{equation}
Similarly, the system entropy does not change after each period, hence:
\begin{equation}
 \bar{\dot{S}}=\bar{\dot{S}}_i+\bar{\dot{S}}_e=0.
 \end{equation}
In combination with the first law, this leads to:
 \begin{equation}
\bar{\dot{S}}_i=-\bar{\dot{S}}_e=-\frac{\bar{\dot{Q}}}{T}=\frac{\bar{\dot{W}}}{T}.
 \end{equation}
 In words, the work performed on the system during each period is dumped, in its entirety, under the form of heat into the reservoir. Prigogine provided  another more revealing description of  this state of
 affairs: we are dealing here with a prototype of a dissipative system. The particle is in a time-periodic nonequilibrium state. The nonequilibrium nature of this state
entails internal irreversible entropy production. The persistence of this nonequilibirum state is only possible because the system imports a compensating negative
entropy flow from the reservoir.

To conclude the analysis, we use the previously derived explicit expression for the heat flow or work flow. One  observes that the 
entropy production (averaged over one period), is quadratic in the amplitude to the driving.  One thus reproduced the  "usual" expression of the entropy production familiar from usual "steady state" linear irreversible thermodynamics:
\begin{equation}\label{epeq}
  \bar{\dot{S}}_i=\frac{\bar{\dot{W}}}{T}=JX\;\;\;\;J=LX\;\;\;\; X=\frac{F_0}{T}.
\end{equation}
The thermodynamic force $X$ is taken to be the  amplitude  $F_0$ of the external driving divided by the reservoir temperature $T$, in agreement with standard linear
irreversible thermodynamics.  
From $\dot{W}=F\dot{Z}$ with $Z$ given by Eq.~(\ref{Zeq}), one finds the following explicit expression for the Onsager coefficient $L$:
\begin{multline}\label{onsL}
L=\frac{2T}{\kappa\tau\tau_2}\int_0^{\tau}dt\int_0^{\infty}dt' e^{-\frac{t'}{2\tau_1}}\frac{\sinh\left({\sqrt{1-4\tau_1/\tau_2}}\frac{t'}{2\tau_1}\right)}{\sqrt{1-4\tau_1/\tau_2}}\\{g}(t)\dot{g}(t-t').
\end{multline}
As is clear from its relation to the non-negative entropy production, this coefficient has to be positive. An explicit proof follows by expressing the periodic forcing in terms of its Fourier series. From
\begin{equation}
F(t)=F_0\sum_{n=0}^\infty \left\{a_{(n,s)}\sin\left(\frac{2\pi nt}{\tau}\right)+a_{(n,c)}\cos\left(\frac{2\pi nt}{\tau}\right)\right\},
\end{equation}
one finds that the Onsager coefficient is given by:
\begin{equation}
L= \frac{T}{\kappa\tau} \sum_{n=0}^\infty\frac{2\pi^2n^2\alpha_2\left(a_{(n,s)}^2+a_{(n,c)}^2\right)}{\left(1-4\pi^2 n^2\alpha_1\alpha_2\right)^2+4\pi^2n^2\alpha_2^2}.
\end{equation}

The above derivation, while appealing in its simplicity, appears to be unexciting in its implications, aside the fact that it can serve as a simple prototype model of a dissipative system. 
From the mathematical point of view, we have merely succeeded in estimating the dissipation per period
in terms of a positive coefficient $L$. To make the connection with an engine, we recall the basic principle of such a construction: it consists of a motor mechanism,
corresponding to an entropy producing process, which drives another "entropy consuming" process, i.e., with a negative entropy production.

\section{Underdamped harmonic duet}
To build a genuine engine, we repeat the above  analysis  in the presence of two external forces, i.e., we set:
\begin{equation}
  F(t)=F_1(t)+F_2(t)\;\;\; F_1(t)=F_1g_1(t)\;\;\;F_2(t)=F_2g_2(t).
\end{equation}
One now distinguishes the work done by each of the forces:
\begin{equation}\label{2feq}
    \dot{W}=\dot{W}_1+\dot{W}_2,\;\;\;\;\dot{W}_1={F}_1\dot{Z},\;\;\dot{W}_2={F}_2\dot{Z}.
\end{equation}
Considering time-periodic modulations with the same period, we note that Eqs.~(\ref{epb}), (\ref{flpeq}) and (\ref{epeq}) remain valid, with the above replacement for the expression of the (total) work. The system can now act like a catalyst in chemistry: it  returns to its original state after each period, having  mediated the exchange of work between the two work sources.  Inserting Eq.~(\ref{2feq}) into Eq.~(\ref{epeq}), one finds (again after averaging over one period):
\begin{equation}
  \bar{\dot{S}}_i=\frac{\bar{\dot{W}}_1+\bar{\dot{W}}_2}{T}= {\bf XLX}={\bf JX}.
\end{equation}
Hence, instead of a single force $X$, flux $J$ and Onsager coefficient $L$, one now has two forces ${\bf X}=(X_1=F_1/T,X_2=F_2/T)$ with corresponding  fluxes ${\bf J}=(J_1,J_2)$ linked by a $2$ by $2$ Onsager matrix ${\bf L}$, ${\bf J}={\bf LX}$.
Furthermore, it is now  possible to extract work $\bar{\dot{W}}_1\leq 0$, i.e.,  the worksource $1$ is  receiving work, provided worksource $2$  "pays for it" by delivering positive work  $\bar{\dot{W}}_2\geq -\bar{\dot{W}}_1 \geq 0$. The efficiency of this "work-to-work-converter" is obviously given by
\begin{equation}\label{effduet}
    \overline{\eta}=-\frac{\bar{\dot{W}}_1}{\bar{\dot{W}}_2}=-\frac{J_1 X_1}{J_2 X_2}=-\frac{L_{11} X_1^2+L_{12} X_1X_2}{L_{21} X_1X_2+L_{22} X_2^2}\leq  1.
\end{equation}
To get the explicit expression of the output power $-\bar{\dot{W}}_1=-J_1 X_1$ and efficiency  $  \overline{\eta}$ in terms of the applied thermodynamic forces ${\bf X}$, we need the expression for the 
Onsager matrix $\bf L$. The latter can be basically copied from the expression Eq.~(\ref{onsL}) following the splitting of the single force into a duet of forces. One finds:
\begin{multline}
L_{ij}=\frac{2T}{\kappa\tau\tau_2}\int_0^{\tau}dt\int_0^{\infty}dt' e^{-\frac{t'}{2\tau_1}}\frac{\sinh\left({\sqrt{1-4\tau_1/\tau_2}}\frac{t'}{2\tau_1}\right)}{\sqrt{1-4\tau_1/\tau_2}}\\\dot{g}_i(t)g_j(t-t').\label{Ons}
\end{multline}
One can again decompose the Onsager coefficients in terms of Fourier components. Setting
\begin{equation}
F_i(t)=F_{i,0}\sum_{n=1}^\infty \left\{a_{(i,n,s)}\sin\left(\frac{2\pi nt}{\tau}\right)+a_{(i,n,c)}\cos\left(\frac{2\pi nt}{\tau}\right)\right\},
\end{equation}
one finds that the Onsager coefficients $L_{ij}$ are given by the following bilinear expression in terms of the Fourier amplitudes ($\sigma,\rho=s,c$ refer to sine and cosine contributions, respectively):
\begin{equation}
L_{ij}=\sum_{\sigma,\rho=\{s,c\}}\sum_{n',n=1}^{\infty}a_{(i,n',\sigma)}L_{(i,n',\sigma),(j,n,\rho)}a_{(j,n,\rho)}.
\end{equation}
with
\begin{eqnarray}
L_{(i,n',\sigma),(j,n,\sigma)}&=&\frac{T}{\kappa\tau} \delta_{n,n'} \frac{2n^2\pi^2\alpha_2a_{(i,n,\sigma)}a_{(j,n,\sigma)}}{\left(1-4\alpha_1\alpha_2n^2\pi^2\right)^2+4\pi^2n^2\alpha_2^2},\nonumber\\
L_{(i,n',s),(j,n,c)}&=&\frac{T}{\kappa\tau} \delta_{n,n'} \frac{n\pi\left(1-4\alpha_1\alpha_2n^2\pi^2\right)}{\left(1-4\alpha_1\alpha_2n^2\pi^2\right)^2+4\pi^2n^2\alpha_2^2},\nonumber\\
L_{(i,n',c),(j,n,s)}&=&-L_{(i,n',s),(j,n,c)}.\label{Lfour}
\end{eqnarray}
We  note that different frequencies do not couple to one another.  This is of course a consequence of the linearity of the underlying dynamics and hence not a deep symmetry principle. 
Furthermore, the matrix consists of symmetric and anti-symmetric parts. This observation is put in the proper perspective 
by considering the Onsager matrix $\tilde{L}_{ij}$ for the time-reversed driving, $\tilde{g}_i(t)=g_i(-t)$:
\begin{eqnarray}
\tilde{L}_{ij}&=&\frac{2T}{\tau\kappa}\int_0^{\tau}dt\int_0^{\infty}dt' e^{-\frac{t'}{2\tau_1}}\frac{\sinh\left(\frac{\sqrt{1-4\tau_1/\tau_2}}{2\tau_1}t'\right)}{\tau_2\sqrt{1-4\tau_1/\tau_2}}\nonumber\\&&\qquad\qquad \qquad\qquad\qquad\qquad\dot{g}_i(-t)g_j(-t+t')\nonumber\\&=&\frac{2T}{\tau\kappa}\int_0^{\tau}dt\int_0^{\infty}dt' e^{-\frac{t'}{2\tau_1}}\frac{\sinh\left(\frac{\sqrt{1-4\tau_1/\tau_2}}{2\tau_1}t'\right)}{\tau_2\sqrt{1-4\tau_1/\tau_2}}\nonumber\\&&\qquad\qquad\qquad\qquad\qquad\qquad\qquad \dot{g}_j(t)g_i(t-t')\nonumber\\&=&L_{ji}.
\end{eqnarray}
In the transition to the second line, we have used  a partial integration with respect to  $t$ and shifted the time-axis of $t$ (using the fact that $g_i(t)$ is time-periodic).  We thus conclude that the Onsager matrix satisfies an Onsager-Casimir symmetry relation, i.e., it is symmetric upon inverting the quantities that are odd under time reversal.
This completes the thermodynamic analysis of the harmonic duet functioning as a work-to-work converter. In the next section we review some of its consequences for the efficiency of the engine.
\begin{figure}
\center
\includegraphics[width=0.45\textwidth]{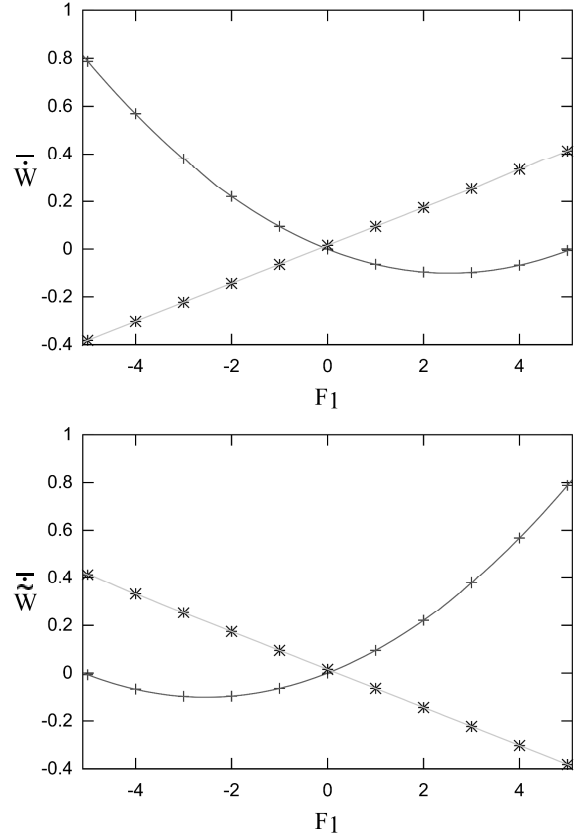} 
\caption{$\bar{\dot{W}}_1$ and $\bar{\dot{W}}_2$ for the Brownian duet with $F_1(t)=F_1\cos(2\pi t/\tau)$, $F_2(t)=F_2\sin(2\pi t/\tau)$, $F_2=1$, $\tau_1=\tau_2=\tau$, for time-forward (top) and time-reversed process (bottom). The diagonal terms $L_{11}$ and $L_{22}$ induce a quadratic dependence in  $F_1$ and a constant contribution for $\bar{\dot{W}}_1$ and $\bar{\dot{W}}_2$, respectively, while the off-diagonal terms give linear contributions.. A straighforward fitting procedure leads to $L_{11}=L_{22}=\tilde{L}_{11}=\tilde{L}_{22}=0.0156$, $L_{12}=-L_{21}=\tilde{L}_{21}=-\tilde{L}_{12}=0.0795$, verifying Onsager-Casimir symmetry.} 
\label{fig3}     
\end{figure}

\section{Efficiency of the  harmonic duet}
\begin{figure}
\center
\includegraphics[width=0.45\textwidth]{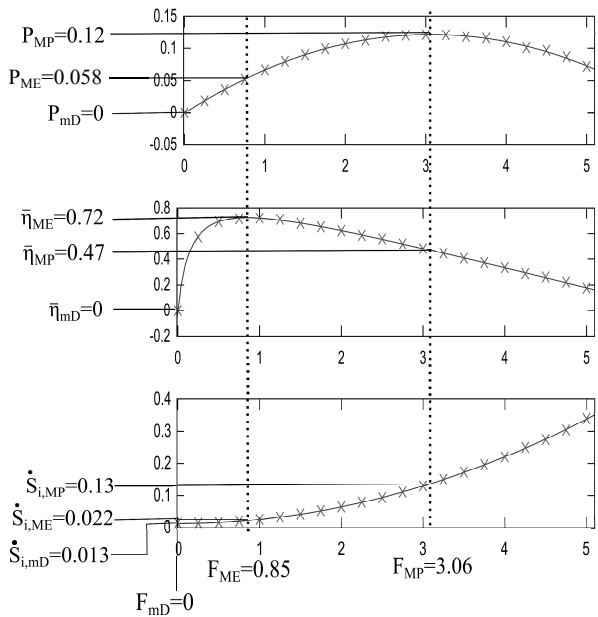} 
\caption{(a) Power, (b) efficiency and (c) dissipation in function of $F_1$, for a Brownian duet, with $F_1(t)=F_1\cos(2\pi t/\tau)$, $F_2(t)=F_2\sin(2\pi t/\tau)$,  $\tau_1=\tau_2=\tau$ and scaling $\tau=\kappa=F_2=1$. One verifies that $P_{ME}/P_{MP}=0.478=1-\bar{\eta}_{ME}^2$, $\bar{\eta}_{ME}/(1+\bar{\eta}_{ME}^2)=0.47=\bar{\eta}_{MP}$, $P_{mD}=0$ and $(1/\bar{\eta}_{MP}-2)P_{MP}=0.013=T\dot{S}_{i,mD}$} 
\label{fig2}     
\end{figure}

The trade-off between the efficiency, power and dissipation of an engine is an important issue, which has been extensively discussed in the literature \cite{van2005thermodynamic,izumida2009onsager,izumida2010onsager,benenti2011thermodynamic,allahverdyan2013carnot,ludovico2016adiabatic,brandner2013strong,shiraishi2015attainability,jiang2014thermodynamic,brandner2015thermodynamics,proesmans2015onsager,raz2016geometric,ponmurugan2016attainability,jiang2016enhancing,lee2016efficiency,PhysRevE.90.062140,ryabov2016maximum,einax2016maximum,proesmans2016linear,shiraishi2016universal,bauer2016optimal,proesmans2016power,cerino2016linear,mukherjee2016speed,yamamoto2016linear,camerin2016fluctuating,campisi2016power,rosas2016onsager,rosas2016onsager2,wang2016unified,johnson2017approaching,gonzalez2017maximum}.  Going back to early work by Moritz von Jacobi on maximizing the output power, an interesting scenario consists in optimizing thermodynamic features with respect to the load force. In the present setting of a time-periodic driving, we will assume that the time-dependence of functions $g_1(t)$ and $g_2(t)$ is specified. We  select an output load amplitude  $F_1$, such that it maximizes  output power or efficiency, or minimizes dissipation. These three different regimes are identified by the subscript notation MP, ME or mD, respectively. Power, efficiency and dissipation  are, via their definitions, linked to each other  in the linear regime as follows:
\begin{equation}
    T\bar{\dot{S}}_i=P\left(\frac{1}{\eta}-1\right)
\end{equation}
A straightforward algebraic calculation shows that the values of these quantities in the   MP, ME and mD regimes are further constrained by the following set of relations \cite{jiang2014thermodynamic,proesmans2016power,bauer2016optimal}:
\begin{eqnarray}
\bar{\eta}_{MP}&=&\frac{P_{MP}}{2P_{MP}-P_{ME}}\bar{\eta}_{ME}\nonumber\\
T\bar{\dot{S}}_{i,mD}&=&\left(\frac{1}{\bar{\eta}_{MP}}-\frac{1}{\bar{\eta}_{ME}^2}-1\right){P_{MP}}+\frac{1}{\bar{\eta}_{ME}^2}P_{ME},\nonumber\\P_{mD}&=&P_{MP}-\frac{1}{\bar{\eta}_{ME}^2}\left(P_{MP}-P_{ME}\right).\label{ped2}
\end{eqnarray}
In the present case,  the Onsager matrix obeys  the following  symmetry relation:
\begin{equation}
    {L}_{12}=\pm{L}_{21}\label{symrem}.
\end{equation}
Under this condition, one can derive the additional result that the power at minimum dissipation vanishes, implying the following simplification of Eqs.~(\ref{ped2}):
\begin{eqnarray}
P_{mD}=0,&&\,\;\,T\dot{S}_{i,mD}=\left(\frac{1}{\bar{\eta}_{MP}}-2\right)P_{MP},\nonumber\\
\frac{P_{ME}}{P_{MP}}=1-\bar{\eta}_{ME}^2,&&\;\;\;\bar{\eta}_{MP}=\frac{\bar{\eta}_{ME}}{1+\bar{\eta}_{ME}^2}.\label{ped}
\end{eqnarray}
An illustrative verification of all these relations is given  in Fig.~\ref{fig2}.

\section{Stochastic dynamics}
The analysis of the previous section can be extended to the study of a periodically  driven Brownian particle in a harmonic potential, by adding a noise term to the equation of motion, Eq.~(\ref{meq}):
\begin{equation}\label{laneq}
    m\ddot{z}(t)=-\gamma \dot{z}(t)-\kappa z(t)+F(t)+\sqrt{2\gamma k_BT}R(t).
\end{equation}
 $R(t)$ is delta correlated noise,
\begin{equation}
\left\langle R(t)\right\rangle=0,\qquad \left\langle R(t)R(t')\right\rangle=\delta(t-t').
\end{equation}
The amplitude of the noise is chosen such that it obeys the fluctuation dissipation relation. 
We are using a lower case notation to stress that the position $z(t)$ of the particle is now a stochastic, fluctuating  quantity.
However, as the above equation is linear, one immediately verifies that $Z(t)=\langle z(t)\rangle$ still obeys Eq.~(\ref{meq}). Similarly one finds that the stochastic power, again using 
the lower case convention to identify the corresponding stochastic quantities:
\begin{equation}
{\dot{w}}_i= F_i (t) \dot{z}(t),
\end{equation}
reduces upon averaging to the previously introduced power  $ \langle {\dot{w}}_i\rangle= \dot{W}_i$.
For this reason, the the linear thermodynamics of the previous section describes the ensemble properties of the above stochastic model. In particular, ensemble averaged response and efficiency are quantified in terms of the aforementioned Onsager-Casimir coefficients.

In the remainder of this paper we however show how other stochastic properties of the model are also linked to the Onsager coefficients.
The first connection is not really novel, as it is an expression of the famous fluctuation dissipation theorem (although here derived in the context of a time-periodic system).
We consider the  fluctuations in the power output:
\begin{eqnarray}
C_{ij}(t)&=&\left\langle {w}_i{w}_j\right\rangle-\left\langle {w}_i\right\rangle\left\langle {w}_j\right\rangle,
\end{eqnarray}
where the notation $w_i$ stands for the change over a time interval $[0,t]$:
\begin{equation}
     w_i=\int^{t}_0dt'\dot{w}_i(t').
\end{equation}
We omit the explicit dependence on $t$ for notational simplicity, whenever it is clear from the context.
In the appendix, we derive  the following  expression for $ C_{ij}(t)$ in the  limit  $t=n\tau$ with the number  $n$ of cycles large. The $\sim$ sign denotes, here and in the sequel, an equality to dominant order in $t$:
\begin{widetext}
\begin{eqnarray}
    C_{ij}(n\tau)\sim \frac{2 k_B T^3 nX_iX_j}{ \kappa  \tau_2}\int^{\tau}_0 dt'\int^{\infty}_0 dt"\left(\dot{g}_i(t'){g}_j(t'-t")+\dot{g}_j(t'){g}_i(t'-t")\right)\frac{e^{-\frac{ t'}{2\tau_1}}\sinh\left(\sqrt{1-4\tau_1/\tau_2}\frac{t'}{2\tau_1}\right)}{\sqrt{1-4\tau_1/\tau_2}}.
\end{eqnarray}
\end{widetext}

Comparison with Eq.~(\ref{Ons}) leads to the fluctuation-dissipation relation:

\begin{equation}
C_{ij}(n\tau)\sim k_B T^2n\tau X_iX_j{\left(L_{ij}+L_{ji}\right)}.\label{fdt}
\end{equation}

\section{Fluctuation theorem}
While we have  reproduced the above fluctuation dissipation relation by an explicit calculation, its validity can be derived directly from the generalization of the second law, describing small scale nonequilibrium systems. To formulate this so-called fluctuation theorem, one needs to define the stochastic analogues of the entropy, energy, heat and work.  These quantities will be denoted by the lower case notation of their macroscopic counterparts. We refer to  \cite{seifert_stochastic_2012,van_den_broeck_ensemble_2014} for an introduction to this stochastic thermodynamics  \cite{schnakenberg1976network,jiu1984stability}, and briefly review the main result that is relevant here. The stochastic system entropy $s$ obeys  a  Prigogine balance equation:
\begin{equation}
    \dot{s}=\dot{s}_i+\dot{s}_e\;\;\;\;\dot{s}_e=\dot{q}/T
\end{equation}
 $\dot{q}$ is the stochastic  heat flux into the system. Contrary to its macroscopic average,
the stochastic entropy production $\dot{s}_i$ need not be positive. In fact, the second law is replaced by a symmetry relation for the probability distribution of this quantity: the fluctuation theorem states that  the probability to have a positive stochastic entropy production rate  is exponentially more likely then to have the corresponding negative entropy production rate  in the process with time-inverted driving \cite{evans1993probability,gallavotti1995dynamical,kurchan1998fluctuation,lebowitz1999gallavotti,harris_fluctuation_2007}:
\begin{equation}
    \frac{p(\Delta{s}_i)}{\tilde{p}(-\Delta{{s}}_i)}\sim e^{\frac{\Delta{{s}}_i }{k_B}},
\end{equation}
with $\Delta{s}_i=\int^{t}_0dt'\dot{s}_i(t') $.
We note in passing that, by multiplying with $\tilde{p}_t(-\Delta{{s}_i})$ and integrating over $\Delta{{s}}_i$, one finds the so-called integral version of the fluctuation theorem, which in turn  implies by Jensen's inequality the usual second law property for the
ensemble average:
\begin{equation}
    \left\langle e^{\frac{{\Delta{s}}_i}{k_B}} \right\rangle=1\rightarrow \langle \Delta{s}_i\rangle \geq 0.
\end{equation}
For the application to the present situation, one needs a "stronger"  fluctuation theorem expressed in terms of the individual fluxes \cite{garcia2010unifying,sinitsyn2011fluctuation}:
\begin{equation}
    \frac{p_t({w}_1,{w}_2)}{\tilde{p}_t(-{w}_1,-{w}_2)}\sim e^{\frac{{w}_1+{w}_2}{k_BT}},\label{fluthe}
\end{equation}
where we have used the fact that the stochastic entropy production in the long time limit is equal to the work input divided by the temperature.
Inserting the Gaussian expression for the work distribution:
\begin{multline}\label{workg}
    p_t({w}_1,{w}_2)=\frac{1}{2\pi \sqrt{\det C}}\\e^{-\frac{1}{2}\sum_{i,j}\left({w}_i-\left\langle {w}_i\right\rangle_t\right)C^{-1}_{ij}\left({w}_j-\left\langle {w}_j\right\rangle_t\right)},
\end{multline}
and an analogous relation for the time-inverted dynamics, one finds in the long time limit that:
\begin{eqnarray}
\frac{2\left({w}_1+w_2\right)t^2}{k_BT}&\sim&-\sum_{i,j}\left(w_i-\left\langle w_i\right\rangle_t\right)C^{-1}_{ij}\left(w_j-\left\langle w_j\right\rangle_t\right)\nonumber\\&&+\sum_{i,j}\left(w_i+\left\langle {\tilde{w}}_i\right\rangle_t\right)\tilde{C}^{-1}_{ij}\left(w_j+\left\langle {\tilde{w}}_j\right\rangle_t\right).\label{fluthe2}\nonumber\\
\end{eqnarray}
Noting that this should hold for any values of the ${{w}_i}$, one has:
\begin{eqnarray}
C(t)&\sim&\tilde{C}(t)\label{cct}\\
C^{-1}(t)\left(\left\langle\bold{{w}}\right\rangle_t+\left\langle\bold{{\tilde{w}}}\right\rangle_t\right)&\sim&\frac{1}{k_BT}\bold{1},\label{cct2}\\
\left\langle\bold{{w}}\right\rangle_t C^{-1}(t)\left\langle\bold{{w}}\right\rangle_t&\sim&
\left\langle\bold{{\tilde{w}}}\right\rangle_t C^{-1}(t)\left\langle\bold{\tilde{{w}}}\right\rangle_t\label{cct3},
\end{eqnarray}
where $\bold{1}=(1,1)$ and where we used the fact that $C$ is by definition symmetric. Plugging Eq.~(\ref{cct2}) into Eq.~(\ref{cct3}) gives,
\begin{equation}
    \bold{1}C(t)\bold{1}\sim2k_BT\bold{1}\left\langle\bold{{w}}\right\rangle_t=2k_BT^2n\tau\bold{X}L\bold{X},
\end{equation}
with $\bold{X}=(X_1,X_2)$ and where we consider $t=n\tau$ with $n$ large in the last equality sign. As this equation should hold for any choice of $\bold{X}$, one reproduces Eq.~(\ref{fdt}), i.e., the fluctuation theorem reproduces the fluctuation-dissipation theorem.

\section{Stochastic efficiency \label{stoef}}
\begin{figure}
\center
\includegraphics[width=0.45\textwidth]{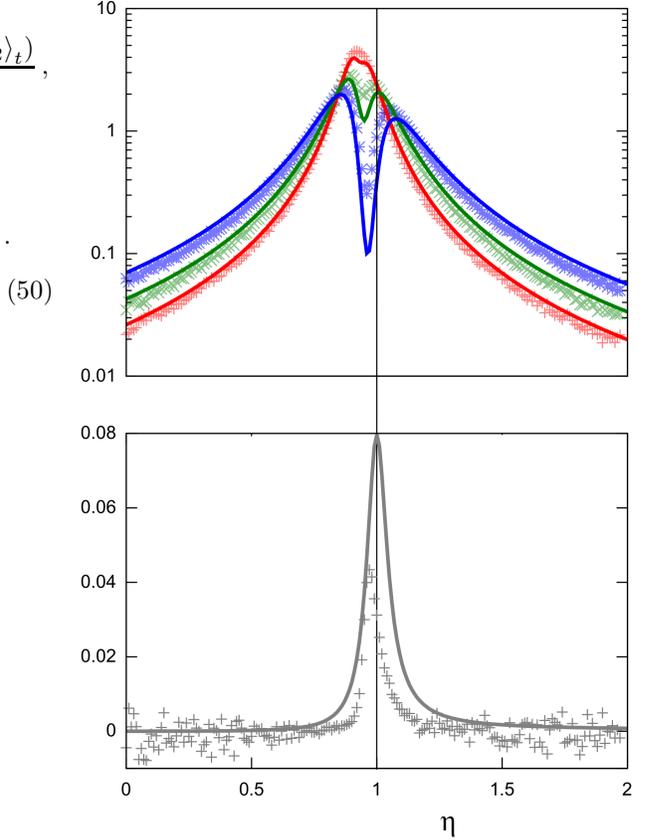} 
\caption{Stochastic efficiency of a Brownian duet with time-symmetric driving: $g_1(t)=10 \cos\left(2\pi t/\tau\right)$ and $g_2(t)=10 \cos\left(2\pi t/\tau\right)+\cos\left(4\pi t/\tau\right)$ with thermodynamic forces $X_1=-X_2=10$, which leads to $\bar{\eta}=0$. Upper panel: probability distribution of the efficiency after $32$ (red), $64$ (green) and $128$ (blue) cycles with analytical results and simulation data. Lower panel: Large deviation function of the efficiency, with analytical results and extrapolation from simulation data (using the extrapolation procedure described in \cite{proesmans_stochastic_2015,Proesmansfcs}).} 
\label{fig4}     
\end{figure}

\begin{figure}
\center
\includegraphics[width=0.45\textwidth]{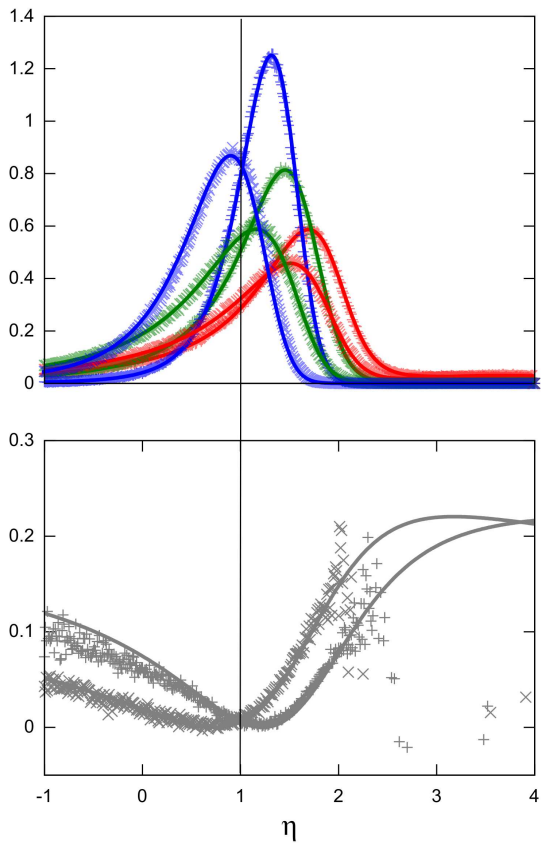} 
\caption{Stochastic efficiency of a Brownian duet with time-asymmetric driving: $g_1(t)=10 \cos\left(2\pi t/\tau\right)$ and $g_2(t)=10 \cos\left(2\pi (t/\tau-0.4)\right)$ with thermodynamic forces $X_1=2$, $X_2=1$. Upper panel: probability distribution of the efficiency after $16$ (red), $32$ (green) and $64$ (blue) cycles with analytical results and simulation data. Lower panel: Large deviation function of the efficiency, with analytical results and extrapolation from simulation data.} 
\label{fig5}     
\end{figure}

A recent discovery in the field of stochastic thermodynamics has to do with the properties of the stochastic efficiency \cite{verley_unlikely_2014,verley_universal_2014,rana_single-particle_2014,gingrich_efficiency_2014,esposito2015efficiency,polettini_efficiency_2015,proesmans2015efficiency,proesmans_stochastic_2015,pilgram2015quantum,Proesmansfcs,jiang2015efficiency,agarwalla2015full,1367-2630-17-9-095005,vroylandt2016efficiency,martinez2016brownian,dinis16,crepieux16,okada2016heat}. The latter is defined as
\begin{equation}
\eta=-\frac{ w_1}{ w_2}.
\end{equation}
One expects that this stochastic quantity will converge to the thermodynamic efficiency $\bar{\eta}$ in the limit of long times $t$. The approach of this limit however holds some surprises, which can be nicely illustrated in the present model.
The probability distribution for the efficiency is given by:
\begin{equation}
p_t(\eta)=\int d  w_1\int d w_2\, p( w_1, w_2)\delta\left(\eta+\frac{  w_1}{  w_2}\right).
\end{equation}
As the probability distribution associated with the work is purely Gaussian, cf.~Eq.(\ref{workg}), the efficiency distribution can be calculated exactly for all times:
\begin{equation}
\begin{aligned}
	p_t(\eta)= \frac{e^{c(\eta)}
		\left[ 2+\frac{\left| a(\eta) \right|}{\sqrt{b(\eta)}} e^{\frac{a(\eta)^2}{2b(\eta)}}\sqrt{2\pi} 
		\text{ erf} \left( \frac{\left| a(\eta) \right|}{\sqrt{2b(\eta)}} \right) \right]}
		{2b(\eta)\pi \sqrt{\det \bold{C}(t)}} \,,
\label{EfficiencyProb}
\end{aligned}
\end{equation}
with
\begin{eqnarray}
	a(\eta) &=&\frac{C_{22}(t) \eta \left<w_1\right>_t-C_{11}(t) \left<w_2\right>_t+C_{12}(t)
		(\left<w_1\right>_t-\eta \left<w_2\right>_t)}{{\det \bold{C}(t)}}\,,\nonumber \\[3pt]
	b(\eta) &=&\frac{C_{11}(t)+2C_{12}(t)\eta+C_{22}(t)\eta^2}{{\det \bold{C}(t)}} \,,\nonumber \\[3pt]
	c(\eta) &=&-\frac{C_{22}(t)\left<w_1\right>_t^2-2C_{12}(t)\left<w_1\right>_t\left<w_2\right>_t-C_{11}(t)
		\left<w_2\right>_t^2}{2 \, {\det \bold{C}(t)}}\,.\nonumber\\
\label{EfficiencyProbCoefs}
\end{eqnarray}
One can straightforwardly check that
\begin{equation}
    p_t(\eta)\sim \eta^{-2},
\end{equation}
for $\eta\rightarrow \pm \infty$. This implies that the moments, and in particular the average and the cumulant generating function, do not exist. While this may seem to be counter-intuitive, one has to realize that the efficiency is not an "additive" quantity, but rather the ratio of "additive" quantities, and therefore has some unusual properties. Furthermore, the macroscopic efficiency is well defined and given by $\bar{\eta}= -\lim_{t \rightarrow \infty} \langle {{w}}_1\rangle/\langle {{w}}_2\rangle$. The properties of the stochastic efficiency are particularly interesting as one approaches this asymptotic limit. The probability distribution for the efficiency converges to a delta function centered at the macroscopic efficiency $\bar{\eta}$, with all other  efficiencies  exponentially unlikely.  More explicitly, this asymptotic behavior is  described
by the so-called large-deviation-function $J(\eta)$ \cite{touchette_large_2009}:
\begin{eqnarray}
J(\eta)=-\lim_{\eta\rightarrow\infty}\frac{1}{t}\ln p_t(\eta).
\end{eqnarray}
By applying this limit to Eq.~(\ref{EfficiencyProb}) and combining it with  the fluctuation-dissipation result, Eq.~(\ref{fdt}), one finds that $J(\eta)$ can be expressed as follows in terms of the Onsager matrix:
\begin{eqnarray}
J(\eta)=\frac{1}{4 k_B}
	\frac{\left( 
		\begin{bmatrix} X_1 & \eta X_2 \end{bmatrix} \boldsymbol{L}
		\begin{bmatrix} X_1 \\ X_2 \end{bmatrix} 
	\right)^2}
		{\begin{bmatrix} X_1 & \eta X_2 \end{bmatrix} \boldsymbol{L}
		\begin{bmatrix} X_1 \\ \eta X_2 \end{bmatrix}} .
\end{eqnarray}
One  verifies the following remarkable properties. $J(1)$ is invariant under a transposition of the Onsager matrix implying $J(1)=\tilde{J}(1)$. Furthermore, $J(\eta)$ has a unique maximum at $\eta=1$ if the Onsager matrix is symmetric, $L_{12}=L_{21}$, which is the case when the driving is time-symmetric. In particular, the probability distribution will intersect at reversible efficiency in the case of time-asymmetric driving, while for time-symmetric protocols a minimum emerges at reversible efficiency in the efficiency distribution. These properties are verified via simulations and analytical calculations in Figs.~\ref{fig4} and \ref{fig5}.

The above long-time properties are in fact generic, as is clear  by deriving them directly from the fluctuation theorem. From
\begin{equation}
\frac{p_t(w_1,w_2)}{\tilde{p}_t(-w_1,-w_2)}\sim e^{\frac{w_1+w_2}{k_BT}},
\end{equation}
one finds that the large deviation function $I({w}_1,{w}_2)$ of the joint work:
\begin{equation}
I({w}_1,{w}_2)=-\lim_{t\rightarrow\infty}\frac{1}{t}\ln p_t(w_1t,w_2t),\label{ftldf}
\end{equation}
obeys the symmetry property:
\begin{equation}
I(w_1,w_2)-\tilde{I}(-w_1,-w_2)=\frac{w_1+w_2}{k_BT},
\end{equation}
with an analogous relation for the time-inverted dynamics. This large deviation function for the work fluxes is related to the large deviation function for the efficiency via the so called contraction principle:
\begin{equation}
J(\eta)=\min_{-w_1/w_2=\eta} I(w_1,w_2)=\min_{\lambda}I(-\eta\lambda,\lambda).\label{contra}
\end{equation}
Note that for reversible efficiency, $\eta=1$, one has $w_1+w_2=0$, and therefore, using Eq.~(\ref{ftldf}),
\begin{equation}
J(1)=\min_{\lambda}I(-\lambda,\lambda)=\min_{\lambda}\tilde{I}(\lambda,-\lambda)=\tilde{J}(1),
\end{equation}
i.e., the large deviation functions for the efficiency of the time-forward and time-reversed process intersect at $\eta=1$. For the time-symmetric case, we note that the minimisation in Eq.~(\ref{contra}) includes $\lambda=0$, and therefore,
\begin{equation}
J(\eta)\leq I(0,0).\label{ineq}
\end{equation}
On the other hand, Eq.~(\ref{ftldf}) implies
\begin{equation}
I(\lambda,-\lambda)+I(-\lambda,\lambda)=0,
\end{equation}
and as $I(w_1,w_2)$ is generally convex, this implies 
\begin{equation}
J(1)=\min_{\lambda}I(\lambda,-\lambda)=I(0,0).
\end{equation}
Combining with Eq.~(\ref{ineq}) implies that $J(1)$ is the maximum of $J(\eta)$, and reversible efficiency becomes the least likely efficiency.\\

\section{Conclusions}
In this work, we have discussed the underdamped  periodically driven (Brownian) duet in the terminology of Ilya Prigogine. Entropy production, Onsager coefficients and Onsager-Casimir symmetry can be easily derived. The analysis provides a pedagogical illustration of  a periodically driven dissipative system, and with a duet of forces of a work-to-work convertor.  With the addition of thermal noise, the model can be analysed in full analytic detail in the context of stochastic thermodynamics.  In particular the connection to the  fluctuation-dissipation relation, to the fluctuation theorem and to universal properties of  stochastic efficiency can be displayed.

\section{Appendix: work correlation function $C_{ij}$}
 The solution of the Langevin equation, Eq.~(\ref{laneq}), for a particle starting at position $z(0)=z_0$ with initial velocity, $v(0)=v_0$ sampled from the periodic steady state distribution Eq.~(\ref{xvdis}), is given by:
\begin{widetext}
\begin{multline} 
z(t)-\left\langle z(t)\right\rangle=\frac{e^{-\frac{\gamma t}{2m}}\left(\left(\sinh\left({\sqrt{1-4\tau_1/\tau_2}}\frac{t}{2\tau_1}\right)+\sqrt{1-4\tau_1/\tau_2}\cosh\left({\sqrt{1-4\tau_1/\tau_2}}\frac{t}{2\tau_1}\right)\right)z_0+2\tau_1\sinh\left({\sqrt{1-4\tau_1/\tau_2}}\frac{t}{2\tau_1}\right)v_0\right)}{\sqrt{1-4\tau_1/\tau_2}}\\+\frac{2F_0}{\beta \kappa\tau_1\tau_2}\int_0^{\infty}dt' e^{-\frac{t'}{2\tau_1}}\frac{\sinh\left({\sqrt{1-4\tau_1/\tau_2}}\frac{t}{2\tau_1}\right)}{\sqrt{1-4\tau_1/\tau_2}}R(t-t').\label{xxcor1}
\end{multline}
To evaluate the probability distribution of position and velocity in the periodic steady state, it is convenient to  start from Kramers equation:
\begin{equation}
    \frac{\partial}{\partial t}p(z,v;t)=-v\frac{\partial}{\partial z}p(z,v;t)+\frac{\partial}{\partial v}(\gamma v p(z,v;t))+\left(\frac{ z}{\tau_1\tau_2}+\frac{F_1g_1(t)+F_2g_2(t)}{m}\right)\frac{\partial}{\partial v}p(z,v;t)+\frac{1 }{\tau_1m\beta}\frac{\partial^2}{\partial v^2}p(z,v;t).
\end{equation}
One verifies by substitution that the solution reads:
\begin{equation}
    p(z,v;t)=\frac{\beta \sqrt{\kappa m}}{2\pi}e^{-\frac{\beta}{2}\left(m(v-\langle{v}\rangle_t)^2+\frac{\kappa}{2}(z-\langle{z}\rangle_t)^2\right)},\label{xvdis}
\end{equation}
where $\langle{z}\rangle_t$ is the solution of Eq.~(\ref{meq}) and $\langle{v}\rangle_t=d\langle{z}\rangle_t/dt$.
Multiplying Eq.~(\ref{xxcor1}) with $z_0$ and averaging with the distribution given in  Eq.~(\ref{xvdis}) leads to:
 \begin{eqnarray}
 \left\langle z(0)z(t)\right\rangle-\left\langle z(0)\right\rangle\left\langle z(t)\right\rangle &=&\frac{2e^{-\frac{\gamma t}{2m}}\left(\left(\sinh\left({\sqrt{1-4\tau_1/\tau_2}}\frac{t}{2\tau_1}\right)+\sqrt{1-4\tau_1/\tau_2}\cosh\left({\sqrt{1-4\tau_1/\tau_2}}\frac{t}{2\tau_1}\right)\right)\right)}{\beta \kappa}\nonumber\\
 &=&\frac{2}{\beta \kappa\tau_2}\int^{t}_0dt'\,\frac{e^{-\frac{ t'}{2\tau_1}}\sinh\left(\sqrt{1-4\tau_1/\tau_2}\frac{t'}{2\tau_1}\right)}{\sqrt{1-4\tau_1/\tau_2}}.\label{xxcor}
 \end{eqnarray}
 Note that the right hand side is invariant under a shift of the time axis, and therefore stationary.

Turning to the work distribution, one writes:
\begin{eqnarray}
C_{ij}(t)&=& {\left\langle \int^{t}_0dt' \dot{w}_i(t')\int^{t}_0dt"\dot{w}_j(t")\right\rangle}-{\left\langle \int^{t}_0dt' \dot{w}_i(t')\right\rangle\left\langle\int^{t}_0dt"\dot{w}_j(t")\right\rangle}\nonumber\\&=&{T^2X_iX_j}\int^t_0 dt'\int^t_0 dt" \dot{g}_i(t')\dot{g}_j(t")\left(\left\langle z(t')z(t") \right\rangle-\left\langle z(t)\right\rangle\left\langle z(t')\right\rangle\right)\nonumber\\&=&{T^2X_iX_j}\int^t_0 dt'\int^{t'}_0 dt" \left(\dot{g}_i(t')\dot{g}_j(t'-t")+\dot{g}_j(t')\dot{g}_i(t'-t")\right)\left(\left\langle z(t")z(0) \right\rangle-\left\langle z(t")\right\rangle\left\langle z(0)\right\rangle\right).
\end{eqnarray}
This result further simplifies after partial integration and  for  $t=n\tau$:
\begin{eqnarray}
C_{ij}(n\tau)&=&{T^2X_iX_j}\int^{n\tau}_0 dt'\int^{t'}_0 dt" \left(\dot{g}_i(t'){g}_j(t'-t")+\dot{g}_j(t'){g}_i(t'-t")\right)\frac{d}{dt"}\left(\left\langle z(t")z(0) \right\rangle-\left\langle z(t")\right\rangle\left\langle z(0)\right\rangle\right)\nonumber\\
&&-{T^2X_iX_j}\int^{n\tau}_0 dt'\int^{t'}_0 dt" \left(\dot{g}_i(t'){g}_j(0)+\dot{g}_j(t'){g}_i(0)\right)\left(\left\langle z(t')z(0) \right\rangle-\left\langle z(t")\right\rangle\left\langle z(0)\right\rangle\right)\nonumber
\\&=&\frac{2T^2X_iX_j}{\beta \kappa \tau_2}\int^{n\tau}_0 dt'\int^{t'}_0 dt"\left(\dot{g}_i(t')\left({g}_j(t'-t")-g_j(0)\right)+\dot{g}_j(t')\left({g}_i(t'-t")-g_i(0)\right)\right)\frac{e^{-\frac{ t'}{2\tau_1}}\sinh\left(\sqrt{1-4\tau_1/\tau_2}\frac{t'}{2\tau_1}\right)}{\sqrt{1-4\tau_1/\tau_2}}.\nonumber\\
\end{eqnarray}

 \end{widetext}
 \begin{acknowledgments}
The computational resources and services used in this work were provided by the VSC (Flemish Supercomputer Center), funded by the Research Foundation - Flanders (FWO) and the Flemish Government department EWI
\end{acknowledgments}

\end{document}